\documentclass[prl,twocolumn,10pt,aps,longbibliography,superscriptaddress]{revtex4-1}

\usepackage{dcolumn}
\usepackage{bm}
\usepackage{amssymb}
\usepackage{amsfonts}
\usepackage{graphicx}
\usepackage{epstopdf}
\usepackage{amsmath}
\usepackage{comment}
\usepackage{soul}
\usepackage{color}

\newcommand{\be}{\begin{equation}}
\newcommand{\ee}{\end{equation}}
\newcommand{\bea}{\begin{eqnarray}}
\newcommand{\eea}{\end{eqnarray}}

\newcommand{\BEA}{\begin{align}}
\newcommand{\EEA}{\end{align}}

\newcommand{\mH}{\mathcal{H}}
\newcommand{\mC}{\mathcal{C}}

\newcommand{\mI}{\mathcal{I}}
\newcommand{\mP}{\mathcal{P}}
\newcommand{\mJ}{\mathcal{J}}
\newcommand{\mB}{\mathcal{B}}
\newcommand{\mO}{\mathcal{O}}
\newcommand{\mQ}{\mathcal{Q}}

\newcommand{\nef}{n_{\textrm{eff}}}

\makeatletter
\newcommand\figcaption{\def\@captype{figure}\caption}
\makeatother

\newcommand{\newcolor}[1]{\textcolor{black}{#1}}

\graphicspath{{figures/}}

\begin{document}


\title{Glassiness and Lack of Equipartition in Random Lasers:\\ the
  common roots of Ergodicity Breaking in Disordered and Non-linear
  Systems.}

\author{Giacomo Gradenigo$^{1,2}$, Fabrizio Antenucci$^{3,2}$,
  Luca Leuzzi$^{1,2}$} \email{luca.leuzzi@cnr.it} \affiliation{$^1$
  NANOTEC CNR, Soft and Living Matter Lab, Roma, Piazzale A. Moro 2,
  I-00185, Roma, Italy \\ $^2$ Dipartimento di Fisica, Universit\`a di
  Roma ``Sapienza,'' Piazzale A. Moro 2, I-00185, Roma, Italy \\ $^3$
  Institut de Physique Th\'eorique, CEA, Universit\'e Paris-Saclay,
  F-91191, Gif-sur-Yvette, France.}

\begin{abstract}

  We present here for the first time a unifying perspective for the
  lack of equipartition in non-linear ordered systems and the low
  temperature phase-space fragmentation in disordered systems. We
  demonstrate that they are just two manifestation of the same
  underlying phenomenon: ergodicity breaking. Inspired by recent
  experiments, suggesting that lasing in optically active disordered
  media is related to an ergodicity-breaking transition, we studied
  numerically a statistical mechanics model for the nonlinearly
  coupled light modes in a disordered medium under external
  pumping. Their collective behavior appears to be akin to the one
  displayed around the ergodicity-breaking transition in glasses, as
  we show measuring the glass order parameter of the
  replica-symmetry-breaking theory. Most remarkably, we also find that
  at the same critical point a breakdown of energy equipartition among
  light modes occurs, the typical signature of ergodicity breaking in
  non-linear systems as the celebrated Fermi-Pasta-Ulam model. The
  crucial ingredient of our system which allows us to find
  equipartition breakdown together with replica symmetry breaking is
  that the amplitudes of light modes are locally unbounded, i.e., they
  are only subject to a global constraint. The physics of random
  lasers appears thus as a unique test-bed to develop under a unifying
  perspective the study of ergodicity breaking in statistical
  disordered systems and non-linear ordered ones.

\end{abstract}

\maketitle

The study of ergodicity-breaking transitions is one among the most
challenging open problems in statistical mechanics. Ergodicity
breaking is typical of a wide variety of models, spanning from the
physics of viscous liquids to the problem of phase transitions in
inference problems. The systematic study of ergodicity breaking in
statistical systems with quenched disorder started with the seminal
work of Parisi in 1979~\cite{Parisi79}, where he put forward his
proposal of an \emph{``Infinite number of Order Parameters''} for the
low temperature phase of spin glasses.  Since then, the framework to
describe the low temperature phase of spin glasses and glasses became
a paradigmatic one: several strong evidences, gathered across the
years, showed that the \emph{``many states scenario''}~\cite{Mezard87}
is the correct one to understand the properties of a wide class of
complex systems.  This notwithstanding, the exact breaking of the
symmetry between \emph{replicas}, which is the formal way to represent
ergodicity breaking in presence of quenched disorder~\cite{Mezard87},
is proven only for \emph{infinite-dimensional} systems. The latter are
models where the interaction network between the microscopic degrees
of freedom has an infinite-dimensional topology, i.e., any degree of
freedom is connected to any other.

It is therefore worth, in order to finally establish the relevance of
exact results on the glass transition, to address the study of
physical systems truly characterized by an infinite-dimensional
interaction network among their degrees of freedom. This is indeed the
motivation to study the non-linear interactions of electromagnetic
field waves inside an optically-active random medium: the complex
amplitudes of a normal mode expansion of these waves form an
infinite-dimensional interaction network where the scenario depicted
by the mean-field theory of spin-glasses should be robust.
Furthermore, as we are going to show, in these systems ergodicity
breaking can be detected in both the ways it is historically known to
take place: not only as the fragmentation of phase space tracked by
non-trivial values of the Parisi's order parameter, but also as the
lack of equipartition between the fundamental degrees of
freedom. \newcolor{This unprecedented evidence is the most remarkable
  finding of the present study on random lasers: it is at the same
  time a consequence of the dilution of non-linear couplings, which
  are reduced in number from $\mO(N^4)$ to $\mO(N^3)$ for the effect
  of a selection rule characteristic of random lasers---but what is
  relevant is dilution \emph{per se}---and a consequence of the fact
  that the complex amplitudes of light modes are not locally bounded,
  unlike Ising, XY or Heisenberg spins.}

The breaking of ergodicity in the form of lack of equipartition was
observed for the first time in the famous Fermi-Pasta-Ulam (FPU)
numerical experiment~\cite{Fermi55}, done in Los Alamos laboratories
in 1954 using one of the first computers worldwide, MANIAC I. The FPU
'experiment' consisted in the numerical study of the relaxation to
equilibrium for a chain of non-linear oscillators with initial energy
put only on few modes: equipartition was never observed up to the
maximum computer time available (at that time). The results were
buried in Los Alamos laboratories until when, in 1965, they were made
public by Stanley Ulam. Immediately after that, the famous paper of
Zabusky and Kruskal~\cite{Zabusky65} on solitons came out,
highlighting the deep connection between the evolution of far from
equilibrium initial conditions in the FPU model and solitary waves
solutions of the Kortweg De-Vries equation. Since then, the
investigation of ergodicity-breaking localized solution in non-linear
systems became a very important research topic on its
own~\cite{Fermi55,Zabusky65,Livi85,Livi87,Cretegny98,KLR04,Chaos05,Benettin13},
unfortunately without any interplay with the study of the ergodicity
breaking \emph{``\`a la Parisi''} in disordered systems.

The aim of the present work is to move the first step towards the
filling of this gap: we present a system where, for the first time,
ergodicity breaking {\color{black}{in disordered systems phase
    transitions}} and the lack of equipartition typical of non-linear
systems are manifest and independently measurable as concomitant
phenomena.

In order to fully appreciate the large scope of our results let us
then take a step backward and provide a more specific introduction to
random lasers in the context of non-linear optics. Several diverse
complex compounds of optically active random scattering media display
lasing under external power pumping above a certain threshold.  In
their generality, they are called random lasers and are currently
investigated in a large number of settings because of their peculiar
properties relevant for applications to bio-sensing, medical
diagnostics, on-chip spectroscopy and optical imaging.  They are easy
to make, chip and robust, can be very small in size, down to the
micron-scale, and exhibit unique features such as low degree of
spatial coherence, lack of directionality and
bio-compatibility~\cite{Cao03,Florescu04b,Anni04,Cao05,LCOW07,vanderMolen07a,Wiersma08,Fallert09,Andreasen11,Leonetti13c,Bachelard14}.

Recent experiments on random lasers provided evidence of particularly
non-trivial correlations between the so-called {\it shot-to-shot}
fluctuations of the emission
spectrum~\cite{Ghofraniha15,Gomes16,Pincheira16, Supratim16,
  Tommasi16,Lopez18,Tommasi18}.  The latter are compatible with an
organization of mode configurations in clusters of states, similar to
the one occurring for the multitude of thermodynamic states composing
the glassy phase in glass formers~\cite{Mezard87}.  Such a
correspondence has been theoretically explained proving the
equivalence between the distribution of the Intensity Fluctuation
Overlaps (IFO) and the distribution of the overlap between states, the
so-called Parisi overlap, the order parameter of the glass
transition~\cite{Antenucci15f}. The analytic proof has been, though,
derived assuming very narrow-band spectra, such that all modes can be
considered at the same
frequency~\cite{Gordon02,Antenucci15a,Antenucci15b}.  This is not the
case, however, for many realistic multimode lasers, both ordered and
random.  There, the four-waves non-linear mixing between
electromagnetic field modes are controlled by a deterministic
selection rule depending on modes frequencies: interactions are
possible only for the quadruplets of modes whose frequencies satisfy
the condition
\be 
|\omega_{i_1}-\omega_{i_2}+\omega_{i_3}-\omega_{i_4}| < \gamma,
\label{eq:selection-rule}
\ee
with $\gamma$ being the typical line-width of a mode. The importance
of such selection rule in multimode random laser has been recently
experimentally demonstrated in~\cite{Antenucci19}. We will call, hereafter, an
interaction network built on the mode-locking selection rule in
Eq.~(\ref{eq:selection-rule}) a \emph{mode-locked} graph. The study of
the repercussions of such selection rule on nonlinear models is,
therefore, a necessary step to understand the fundamental mechanisms
at the ground of the fascinating phenomenon of lasing in random media.

\begin{figure*}[t!]
  \includegraphics[width=0.9\columnwidth,trim={3.0cm 3.8cm 0.9cm 11.1cm},clip]{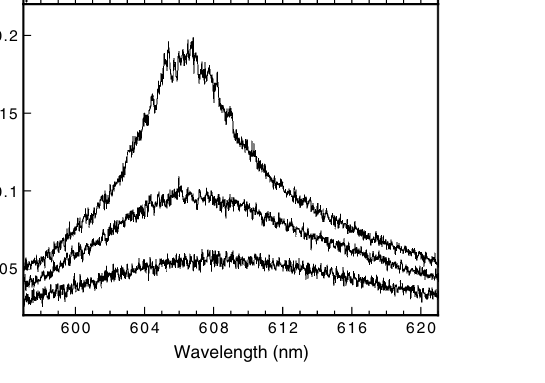}
  \put(-200,120){\LARGE {\bf a)}}
  \includegraphics[width=0.9\columnwidth]{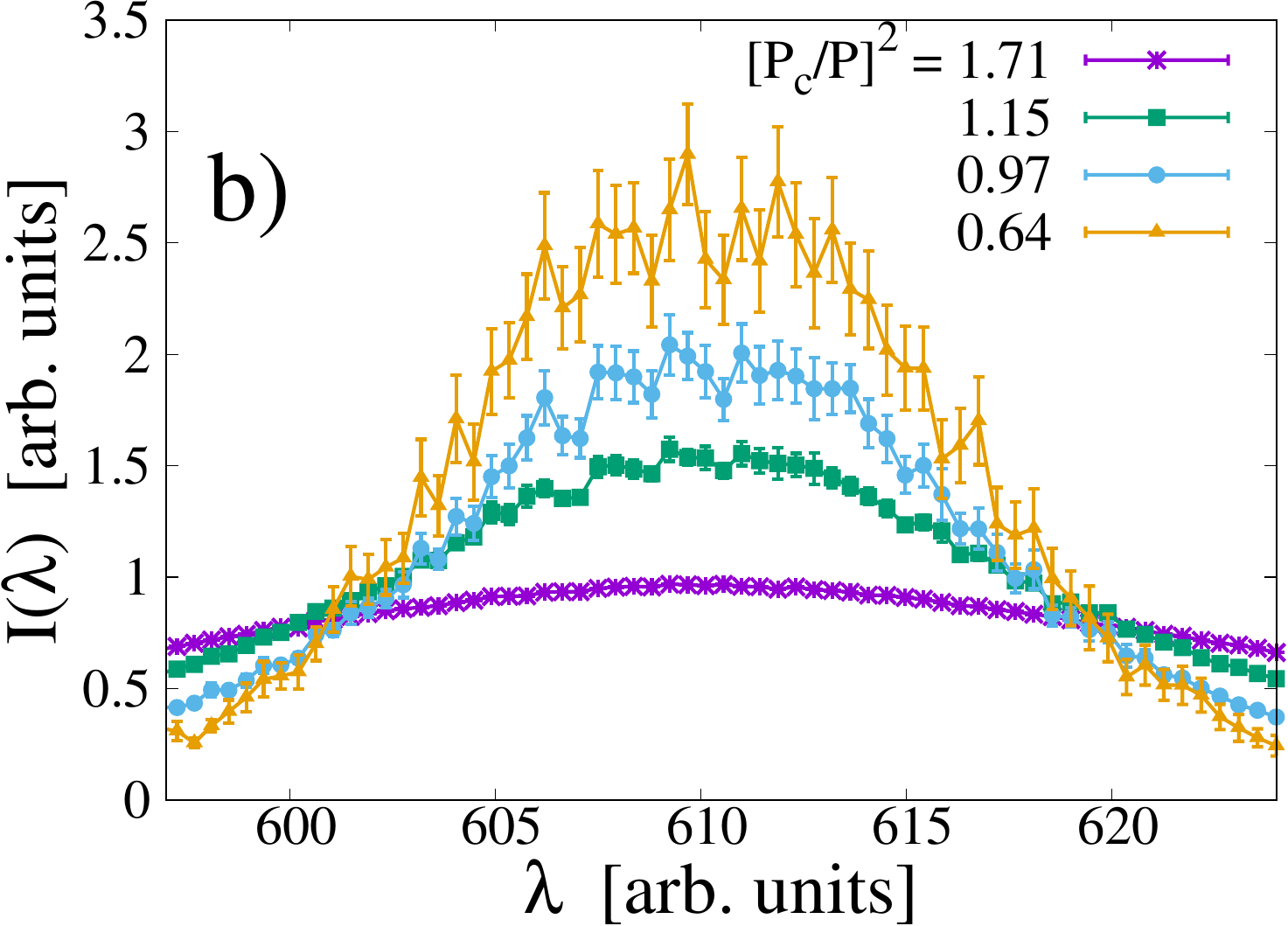}
  \includegraphics[width=1.2\columnwidth]{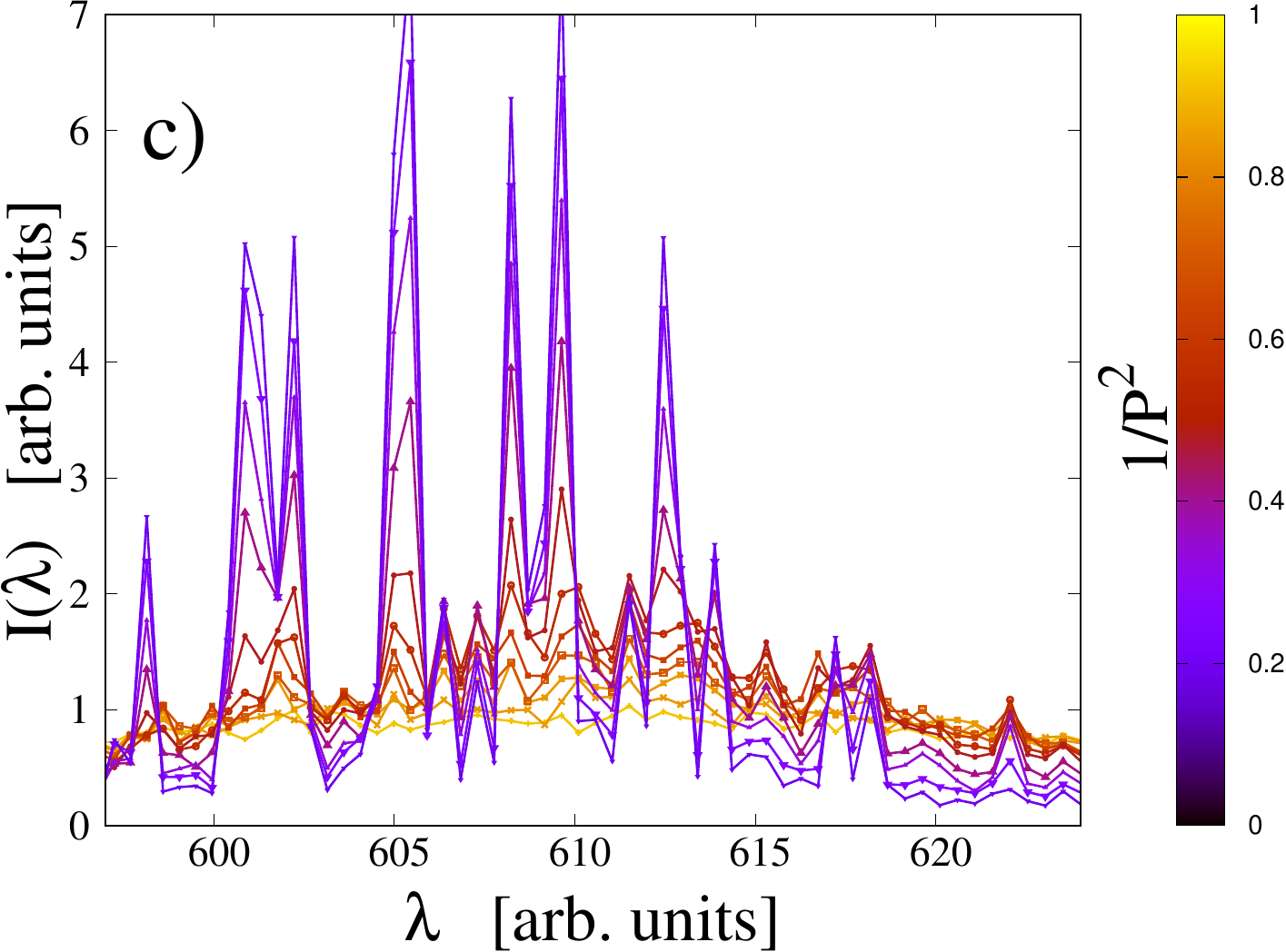}
  \caption{{\it \bf Panel a)}: Emission spectrum of ZnO nanoparticles
    in a rhodamine $640$ dye solution. From bottom to top: increasing
    injected optical power $\mP$. The emission spectrum of rhodamine
    shown here is a reproduction of the original Fig. 9 appearing in
    [H. Cao, {\it Waves Random Media} {\bf 13}, R1 (2003)]. {\it \bf
      Panel b)}: Intensity spectrum $I_\lambda = \langle A_\lambda^2
    \rangle$ as a function of the wavelength $\lambda$ averaged over
    many instances of the quenched randomness, numerical simulations
    with $N=64$ degrees of freedom. The wavelength $\lambda=2\pi/k$ is
    expressed here in arbitrary units ([a.u.]). Notice the narrowing
    of the spectrum at larger values of the pumping rate $\mP$. {\it
      \bf Panel c)}: Intensity spectrum $I_\lambda$ as a function of
    the wavelength $\lambda$ for a single instance of the quenched
    randomness, numerical simulations with $N=64$ degrees of
    freedom. Color code of the curves: $\mP$ increases from bright
    (yellow) to dark (purple). Notice the crossover from a smooth,
    almost equipartited, spectrum at low $\mP$ to a disordered pattern
    of isolated peaks at high $\mP$.}
  \label{fig1}
\end{figure*}

As a first step of this work, we reproduce in numerical simulations
the narrowing of the emission spectrum across the lasing threshold
found in experiments, see ~\cite{Cao03r} and Fig.~\ref{fig1}. We,
further, demonstrate that the narrowing of the spectrum takes place
concomitantly with an ergodicity-breaking transition, the glass
transition, known to occur in random lasers~\cite{Antenucci15a,
  Antenucci15b, Antenucci15f,Ghofraniha15}.  We put in evidence two
salient features of the {\it glassy} phase of light: 1) the breaking
of equipartition between the fundamental degrees of freedom of the
system [Fig.\ref{fig1}]; 2) the rise of non-trivial \emph{glassy}
correlations among different emission events, landmark of the phase
space shrinking typical of an ergodicity-breaking transition. The
discovery that the breaking of equipartition in our model is just
another facet of this ergodicity-breaking does not only provide a
strong connection between the physics of non-linear and of disordered
systems, but it is important also on its own as, from our analysis, it
acquires the status of a \emph{new} tool to detect glassy phases. We
will explain below how breaking of equipartition is revealed by a
spectral analysis which is technically much more easy to perform than
the study of the distribution of the state overlap, i.e., the order
parameter for many state disordered systems.

{\color{black}{For the sake of clarity we underline that the $P(q)$
    distribution that we find is not the full replica symmetry
    breaking one, describing, e.g., the prototypical spin-glass
    Sherrington-Kirkpatrick model. There the probability distribution
    of the overlap $P(q)$ has a nontrivial shape with a continuous
    $q$-support in the thermodynamic limit $N\to \infty$. Here the
    $P(q)$ is expected to be bimodal in the thermodynamic limit,
    though we will see that at finite sizes it appears to be a smooth
    function of $q$ due to strong finite size effects. }}

One last point is important to discuss in the introduction: the lack
of equipartition has never been observed so far in disordered glassy
systems. The reader might, in fact, be quite suspicious that after 40
years of the \emph{discovery} of replica symmetry breaking such a
fundamental phenomenon was never observed. The crucial point relies on
the kind of variables and on the kind of network used in our
model. Here we consider a model with continuous and locally unbounded
variables---only one global constraint keeps the energy bounded---and
the structure of the interaction network is dense but not complete,
therefore inhomogeneous. A specific combination of ingredients that
for diverse reasons has never been studied in detail before. Let us
shortly explain why. In the statistical physics framework complete
graphs with continuous variables, pertaining to the set of $p$-spin
models~\cite{Kirkpatrick87b,Crisanti95}, are usually considered as
analytically solvable approximations to models with discrete Ising
spins. It is the completeness of the graph, i.e., the fact that in
presence of $p$-body interactions \emph{all} the independent
$p$-upletes of degrees of freedom are considered, and hence its
homogeneity, that makes an exact solution
available~\cite{Mezard87,Kirkpatrick87b,Crisanti95}. While always
considered for analytic computations, models on complete graphs are
vice-versa usually not studied in numerical simulations, because of
their high computational cost. On the contrary, numerical simulations
are the main investigation tool for finite-dimensional interaction
networks.  But in the case of a finite-dimensional topology continuous
and unbounded variables were never studied due to the problem of the
\emph{power condensation catastrophe}~\cite{Antenucci15c,
  Antenucci15d}. Models studied extensively on finite-dimensional
topologies are the XY model or Heisenberg spins, where, due to the
fixed modulus of local variables, is not possible \emph{by
  construction} to observe breaking of equipartition. Hence, also from
this respect, our study is completely original.\\

In conclusion, the dense yet non-complete interaction network studied
in the present work, i.e., the \emph{mode-locked} graph, has all the
ingredients to see some new physics: the structure of the graph is not
completely homogeneous and the local variables can change their
magnitude, a lucky combination which allowed us to probe the
non-trivial local fluctuations which give rise to the so far
unobserved breaking of equipartition at the glass transition. Most
remarkably, the structure of the interaction network studied here is
not a \emph{build-on-purpose} one, but is the one produced by the
matching condition among mode frequencies in an optical system with
nonlinear polarization in a random medium under external energy
pumping~\cite{SargentIII78}.\\

\section{Model and Observables}
\label{model-and-observables}

The disordered model studied here is the {\it mode-locked} $4$-phasor
model, characterized by a deterministic selection rule on the
interacting quadruplets of modes which emerges naturally from
the study of mode dynamics in the stationary
regime~\cite{Gordon02,Antenucci15a,Antenucci15b,Antenucci15e,Antenucci16}. The
dynamic variables of the system are the complex {\em slow } amplitudes
$a_k(t) = A_k(t)~e^{i\phi_k(t)}$ of the electromagnetic field
expansion in normal modes
\be
\nonumber
{\bf E}({\bf r},t) = \sum_{k=1}^N
a_k(t)~e^{i\omega_k t}~{\bf E}_k({\bf r})+\textrm{c.c.}.
\ee
Being $A_k=|a_k|\in \mathbb{R}^+$ and $\phi_k=\arg a_k \in [0,2\pi]$,
the dynamics of the stationary regime can be shown to be a stochastic
potential dynamics whose Hamiltonian reads
~\cite{Angelani06a,Antenucci15a}:
\bea && \mH[{\bf a}] = - \sum^{\rm FMC}_{i_1<i_2}g_{i_1i_2}~\bar
a_{i_1}a_{i_2} \nonumber \\ && +\sum_{i_1<i_2<i_3<i_4}^{\rm FMC}
J_{\vec\i}~\bar a_{i_1}a_{i_2}a_{i_3}\bar a_{i_4} + \mbox{ c.c.}
\nonumber \\ && = - \sum_{i_1<i_2}^{\rm FMC}
g_{i_1i_2}~A_{i_1}A_{i_2}\cos(\phi_{i_1}-\phi_{i_2}) \\ &&
- \hspace*{-.3cm}\sum_{i_1<i_2<i_3<i_4}^{\rm FMC} \hspace*{-.3cm}
J_{\vec\i}~A_{i_1}A_{i_2}A_{i_3}A_{i_4}\cos(\phi_{i_1}-\phi_{i_2}+\phi_{i_3}-\phi_{i_4}),
\nonumber \\ &&\qquad\qquad\vec \i \equiv i_1,i_2,i_3,i_4
\eea

An important ingredient of the model is then the implementation of
gain saturation, formally rephrased into a global constraint on the
intensity \cite{Gordon02,Antenucci16}:
\be
\epsilon N = \sum_{k=1}^N A_k^2 \label{eq:sphericalc}
\ee
The diagonal part of the $g_{ii}$ effective linear coupling is the net
gain profile, that plays a role mainly below the lasing threshold.
{The off-diagonal terms can interpreted as effective interactions
  between radiant modes \cite{Hackenbroich03,Viviescas03}, though only
  modes of very similar frequency can display a coupling. They are all
  zero when the frequencies are well distinct, so that the frequency
  matching condition of the linear term is never satisfied unless
  different modes have overlapping frequency.  While this is generally
  true for standard high quality-factor multi-mode mode-locked lasers
  \cite{Haus00}, in {\em random} media a significant frequency overlap
  between modes can occur and off-diagonal linear contributions are
  actually important in the fluorescence regime.

To express the interactions in the slow amplitude mode basis actually
used in the dynamics is, actually, a complicated problem. Actually the
nature and behavior of modes in random media is still the subject of
active research~\cite{Andreasen11}.  In some cases the solution can be
found using some self-consistent procedures, starting from the
solution obtained without the non-linear
coupling~\cite{Tureci06,Tureci08, Tureci09, Rotter14}.  In particular,
when the non-linear term is entirely neglected, a possible (though not
unique) solution is the one that diagonalizes the linear interaction.
When the lasing threshold is overcome, however, the non-linear term
becomes non-perturbatively relevant and the diagonalization of the
linear term does not correspond to a slow amplitude basis, anymore, in
the most general case of lasing in random media.  }

Our theory works for any basis, as far as it is well defined and
complete, and it focuses on the onset lo the lasing phase, and on the
properties of the modes in this regime.  For simplicity, according to
this focus, we make the working choice of a constant gain profile
$g_{ii}= g$, $\forall \, i$, in the whole wave-length band of the
random laser and zero off-diagonal $g$ contributions.  Together with
the global constraint Eq.~(\ref{eq:sphericalc}) this implies the
following form for the Hamiltonian

\bea
&& \mH[{\bf a}] =  -  g \,\epsilon N
\label{eq:hamiltonian}
\\
 && - \hspace*{-.3cm}\sum_{i_1<i_2<i_3<i_4}^{\rm FMC} \hspace*{-.3cm} J_{\vec\i}~A_{i_1}A_{i_2}A_{i_3}A_{i_4}\cos(\phi_{i_1}-\phi_{i_2}+\phi_{i_3}-\phi_{i_4}), \nonumber 
\eea

The sum termed FMC is generated by choosing the non-linear interactions
according to a selection rule which depends on mode frequencies, the
so-called Frequency Matching Condition (FMC)~\cite{Antenucci15d}:
\be
|\omega_{i_2}-\omega_{i_1}+\omega_{i_3}-\omega_{i_4}| \lesssim \gamma.
\label{eq:fmc}
\ee 
Here $\gamma$ is the typical line-width and we explicitly wrote the
permutation that can satisfy the constraint in the ordered sum over
$i_1<i_2<i_3<i_4$. The FMC constraint introduces in the topology of
the interaction network inhomogeneities such that standard
{\color{black}{\em fully connected} }mean-field approximations used to
solve the thermodynamics of disordered systems~\cite{Mezard87}
{\color{black}{cannot be applied}}. Let us consider for instance the
simple case of a linear dispersion relation with equispaced angular
frequencies, $\omega_j = \omega_0 + j~\delta$, with $\delta \ll
\gamma$ and $i=1,\ldots,N$. Eq.~(\ref{eq:fmc}) very simply reads as
the constraint
\begin{eqnarray}
|i_1-i_2+i_3-i_4|=0
\label{eq:FMC_indici}
\end{eqnarray}
on the summation indices in Eq.~(\ref{eq:hamiltonian}), diluting by an
order $N$ the number of interactions from the fully connected case,
$N(N-1)(N-2)(N-3)/24$~\cite{Marruzzo18}.  We call the resulting
interaction network the {\it mode-locked} graph.

The values of the coefficients $J_{\vec \i = \{i_1,i_2,i_3,i_4\}}$ in
Eq.~(\ref{eq:hamiltonian}) are linked to the electro-magnetic normal
modes and the nonlinear optical susceptibility as
\begin{widetext}
\begin{equation}
J_{i_1,i_2,i_3,i_4} \propto \int d\bm r~ \sum_{\vec\nu}^{x,y,z} E^{(\nu_1)}_{i_1}(\bm r) E^{(\nu_2)}_{i_3}(\bm r)E^{(\nu_3)}_{i_3}(\bm r)E^{(\nu_4)}_{i_4}(\bm r) 
 ~\chi^{(3)}_{(\vec\nu)}(\bm r|  \omega_{i_1},\omega_{i_2}, \omega_{i_3}, \omega_{i_4})
 \label{eq:coupling}
\end{equation}
\end{widetext}
In comparison to the {\em slow} amplitude dynamics these couplings are
even slower, so that they can be taken as quenched.  The spatial
distribution of the normal modes $\bm E(\bm r)$ and the susceptibility
of $\chi^{(3)}(\bm r)$ are not exactly known in realistic random
lasers, nevertheless they are expected to be largely inhomogeneous.
Here we then consider the interaction coefficients
$J_{i_1,i_2,i_3,i_4}$ to be Gaussian distributed random variables as
the statistical behavior of the system would not depend on the
specific form for large $N$.

Despite the fact that energy is continuously injected and dissipated
within a random laser, according to~\cite{Gordon02,
  Angelani06a,Antenucci15a,Antenucci16} one can assume an effective
equilibrium distribution for the amplitudes:
\be
P(a_1,\ldots,a_k) = e^{-\beta \mH[{\bf a}]}~\delta\left( \epsilon N - \sum_{k=1}^N A_k^2\right),
\label{eq:boltzmann}
\ee
where $\beta$ is some effective inverse temperature
{\color{black}{(related to the rate of spontaneous emission of
    photons)}} and $\epsilon$ measures the optical power per mode
available to the system. Rescaling $A_k \to A_k/\sqrt{\epsilon}$ in
Eq.~(\ref{eq:boltzmann}), the new variables are constrained on the
same hypersphere at the cost of a rescaling of the effective
temperature as
\be
\beta\to\beta\epsilon^2=\mP^2
\label{eq:Peff}
\ee
where $\mP$ is the so-called  {\it pumping rate} parameter.
In these rescaled variables the Boltzmann weight reads
\be
\nonumber
\rho[\{\bm a\}] = \frac{1}{Z}\exp\{- \mP^2 \mH[{\bf a}]\} \ ,
\ee
making explicit the role of the pumping as effective heat bath for the
stationary regimes of the lasing  medium
\footnote{We have investigated how the system behaves varying the
  pumping rate $\mP$. According to Eq.~(\ref{eq:boltzmann}) and
  Eq.~(\ref{eq:Peff}), in numerical simulations it is identical to fix
  the constraint $\epsilon$ and change the effective temperature
  $T=\beta^{-1}$ or work at fixed temperature varying the value of
  $\epsilon$. We have done our simulations varying the temperature
  $T$, in order to leave a clear term of comparison with the
  literature on glassy systems (see Methods for the numerical
  algorithm) but we will often discuss our results in terms of pumping
  rate $\mP$. The reader has just to bear always in mind that $\mP
  \sim 1/\sqrt{T}$.}.


\begin{figure*}[t!]
  \includegraphics[width=1.9\columnwidth]{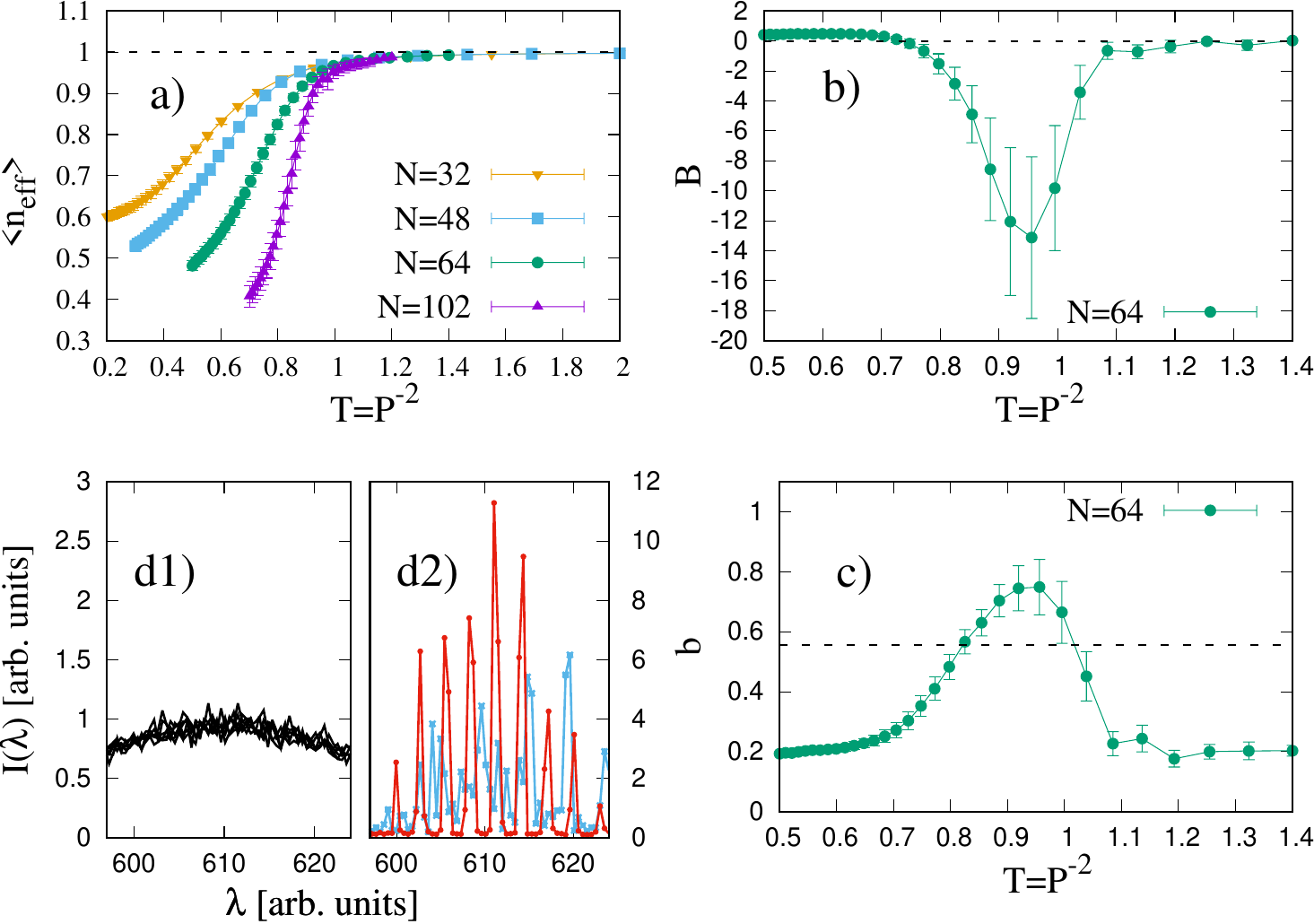}
  \caption{ {\bf Panel a)}: Effective number of degrees of freedom
    $\overline{n}_{\text{eff}}$ (disorder average) as a function of
    the inverse (squared) pumping rate $\mP^{-2}=T$, corresponding to
    the temperature in the numerical
    algorithm. $\overline{n}_{\text{eff}} \approx 1$ signals
    equipartition, $\overline{n}_{\text{eff}} < 1 $ lack of
    equipartition. {\bf Panel b)}: Binder parameter $\mB$ for the
    probability distribution of $\nef$, measured as a function of
    $\mP^{-2}$. {\bf Panel c)}: Bimodality parameter $b$ for $P(\nef)$
    measured as a function of $\mP^{-2}$. {\bf Panel d1) \& d2)}:
    Spectra obtained from different instances of the disorder are
    shown for the equipartited phase, $(\mP_c/\mP)^2 \approx 2$, in
    ${\it \bf d1}$, and the non-equipartited phase,
    $(\mP_c/\mP)^2\approx 0.5 $, in ${\it \bf d2}$.}
\label{fig2}
\end{figure*}


\section{Numerical algorithm}
\label{numerical algorithm}

We have studied systems with $N$ complex variables $a_k =
|A_k|e^{i\phi_k}$ interacting with Hamiltonian in
Eq.~(\ref{eq:hamiltonian}). 

\subsubsection{Exchange Monte Carlo}

The sampling of the probability distribution in
Eq.~(\ref{eq:boltzmann}) was done by means of a Parallel Tempering
Monte Carlo algorithm (PT)~\cite{Hukushima96}. In the PT algorithm one
runs $K$ simulations with local Metropolis dynamics for identical
replicas of the same system, i.e. with the same quenched disorder
$J_{\vec{\i}}$. {\color{black}{Then exchanges of configurations
    between heat baths at nearby temperature are proposed, provided
    that, eventually, in the thermalized systems for each one of the
    $K$ copies the configurations are sampled according to the
    equilibrium Gibbs distribution $e^{-\beta_i E}$.}}

We have run $K$ independent simulations at temperatures
$T_i=\beta_i^{-1}$, from $T_0=T_{\textrm{min}}$ to
$T_{N_{PT}}=T_{\textrm{max}}$. Each $64$ steps of the local Metropolis
algorithm an exchange of configurations among simulations running at
neighbouring temperatures is proposed. That is, if the Metropolis
algorithm for ${\bf a}_i$ runs with $\beta_i$ and that of ${\bf a}_j$
with $\beta_j$ one tries the following attempt: $\lbrace ({\bf
  a}_i,\beta_i); ({\bf a}_j,\beta_j) \rbrace \Longrightarrow ({\bf
  a}_j,\beta_i); ({\bf a}_i,\beta_j)$, which is accepted with
probability $p_{\textrm{swap}}$:
\be
p_{\textrm{swap}} = \textrm{min}\left[ 1, e^{-(\beta_i-\beta_j)(\mH[{\bf a}_j] -\beta_j\mH[{\bf a}_i])}
    \right]
\ee
The replica exchange update is proposed sequentially for all pairs of
neighbouring temperatures $\beta_i$ and $\beta_{i+1}$. For all
simulations the $N_{PT}$ temperatures where taken with a linear
spacing in $\beta$, i.e. $\beta_{i+1} = \beta_i +\Delta\beta$.

\subsubsection{The Monte Carlo update under spherical constraint}

In the local Metropolis algorithm the configuration of complex
``spins'' $(a_1,\ldots,a_N)$ is updated with the requirement of
keeping $\sum_k |a_k|^2=\textrm{const}$. In order to fulfill such
constraint and the detailed balance each update is realized choosing
at random two spins $a_i = A_i e^{i \phi_i}$ and $a_j = A_j e^{i
  \phi_j}$ and proposing an update to randomly chosen values of $a_i'$
and $a_j'$ such that:
\be
|a_i|^2+|a_j|^2 = |a_i'|^2+|a_j'|^2 = C^2 \ .
\ee
This is simply achieved by extracting three random numbers $\phi_i'$,
$\phi_j'$ and $\theta$ with uniform probability in the interval
$[0,2\pi]$ and proposing the four simultaneous updates $\phi_i\rightarrow
\phi_i'$, $\phi_j\rightarrow \phi_j'$, $A_i \rightarrow A_i'=C \cos(\theta)$ and
$A_j \rightarrow A_j'= C \sin(\theta)$. We have
simulated systems at four different sizes: $N=32,48,64,102$.   For the
parallel tempering we used $N_{\rm PT}=32$ temperatures for all sizes. 

\subsubsection{Parallel computation on GPU}

The update of the variables $A_i$ must be done sequentially, since the
interaction network is dense and there is no way to partition the
variables in subsets which can be updated independently in parallel.
We have, instead, implemented parallelization on GPU graphic cards in
two ways: (i) the update energy shift and (ii) the PT replicas
dynamics between configuration exchanges.

In order to accept or reject the update of two spins $a_i$ and $a_j$
one has to compute the energy update on $N_4^{(i,j)}=\mO(N^2)$
quadruplettes, i. e.,
\begin{eqnarray}
\Delta E = \sum_{\kappa=1}^{N_4^{(i,j)}} \Delta E_\kappa
\end{eqnarray}

 The calculation of each $\Delta E_\kappa$ is realized in parallel on
 a distinct kernel on GPU.

Moreover, the execution of the $K$ evolutions at different temperature
of the parallel tempering is quite naturally implemented in parallel
on the GPU. We have run simulations on two type of graphic cards: GTX
$680$, and Tesla K$20$, achieving an overall speedup, with respect to
the sequential code on CPU, up to a factor $8$.
\begin{table}[]
\centering
\begin{tabular}{lrrrrrr}
\hline \hline
$N$ & $N_4$ & $T_{\text{min}}$ & $T_{\text{max}}$ & $N_{\text{PT}}$  & $N_{\text{sweep}}$ & $N_{\text{sample}}$ \\
 \hline 
32 & $2^{11}$  & 0.2 & 2.0 & 32 & $2^{20}$ & 100 \\
48 & $2^{13}$  & 0.3 & 2.0 & 32 & $2^{20}$ & 100 \\
64 & $2^{14}$  & 0.5 & 1.4 & 32 & $2^{20}$ & 100 \\
102 & $2^{16}$ & 0.7 & 1.2 & 32 & $2^{21}$ & 100 \\
\hline \hline
\end{tabular}
\caption{Details for the simulations for different system sizes $N$. The number of quadruplets involved in the Hamiltonian, $N_4$, is always a power of $2$.}
\label{tab1}
\end{table}

\section{Breaking of equipartition}
\label{breaking-of-equipartition}

Let us first compare the dependence on the pumping rate $\mP$ of the
emission spectrum measured in experiments (panel a. of
Fig.~\ref{fig1}, data taken from \cite{Cao98}) with the emission
spectrum in our numerical simulations averaged over the random
coupling distribution (panel b.): the increase of $\mP$ is accompanied
by the same kind of narrowing.  {\color{black}{We stress that the
    symmetry of the spectral distribution around the central
    wave-length is due to the fact that the gain is taken as constant
    in this case, cf. Eq. (\ref{eq:hamiltonian}), whereas in reality a
    material dependent, non-constant, gain curve $g(\lambda)$ is
    present.}}

The equipartition breaking obtained by raising $\mP$ is, then, even
clearer if we consider the behaviour for a single instance of disorder
rather than its average over many realizations, see panel $c)$ of
Fig.~\ref{fig1}. From a smooth profile at low $\mP$ a pattern of
random sharp peaks emerges in the spectrum at high $\mP$. {\color{black}
  Let us point out that the spectrum shown in panel c) of
  Fig.~\ref{fig1} represents an average over a large time window (see
  Appendix), in agreement with the experimental nature of the emission
  spectrum, which is a time-integrated signal.}

From the literature on non-linear systems~\cite{Livi85,Cretegny98} we
know that the degree of equipartition in the spectrum can be measured
by the so-called \emph{spectral entropy}:
\begin{eqnarray}
S_{\textrm{sp}}&=& -\sum_{i=1}^N \hat{\mI}_k\log(\hat{\mI}_k),
\label{eq:spectral-ent}
\\
\hat{\mI}_k &=&\frac{ \langle A_k^2
\rangle}{\left(\sum_{k=1}^N \langle A_k^2\rangle\right)}
\end{eqnarray}
where $\hat{\mI}_k$ is the normalized thermal average of the intensity
for given wave number $k$. From the
spectral entropy one then defines the {\em effective number of
  degrees of freedom}~\cite{Livi85,Cretegny98}:
\be
n_{\text{eff}} = \frac{\exp\left( S_{\text{sp}}\right)}{N},
\label{eq:neff}
\ee
where $n_{\text{eff}}=1$ for perfect equipartition and
$n_{\text{eff}}=1/N$ when the total energy is concentrated in a single
mode~\cite{Livi85,Cretegny98}. For our system the behaviour of its
average, $\overline{n}_{\textrm{eff}}$, as a function of the pumping
rate $\mP$, is reported in Fig.~\ref{fig2}, panel $a)$ [data are
  plotted as a function of $\mP^{-2}=T$].  At low pumping rate we find
$\overline{n}_{\textrm{eff}}\approx 1$, which signals a good degree of
equipartition for all the sizes $N$ studied. On the contrary, by
increasing $\mP$ we find that $\overline{n}_{\textrm{eff}}$ rapidly
decreases and the decrease depends on $N$. {\color{black}{We identify
    as the pumping value at which $\overline{n}_{\textrm{eff}}$ starts
    decreasing below $1$ with a }} size-dependent lasing threshold
$\mP_{c}(N)= \sqrt{T_c(N)}$. The figure clearly shows that the larger
the $N$, the steeper the $\overline{n}_{\textrm{eff}}$ decrease.

A stronger indication that we are dealing with a first-order
transition with respect to $\overline{n}_{\textrm{eff}}$ comes from
the study of the Binder, $\mathcal{B}$, and the bimodality, $b$,
parameters of the distribution $P(n_{\text{eff}})$, displayed in
panels $b)$ and $c)$ of Fig.~\ref{fig2}. The Binder parameter, which
is simply defined as the ratio between the fourth and second moment of
the distribution $P(n_{\text{eff}})$,
\be
\mathcal{B} = \frac{1}{2}\left( 3 - \frac{\overline{(\Delta n_{\text{eff}})^4}}{\left[\overline{(\Delta n_{\text{eff}})^2}\right]^2}\right),
\label{eq:Binder}
\ee
with $\Delta n_{\text{eff}} =
n_{\text{eff}}-\overline{n}_{\text{eff}}$, is a measure of the
deviation from Gaussianity. The order parameter of a first order
transition is typically characterized by a Gaussian distribution,
which is centered around two different valued depending on whether its
is probed well below or well above the transition temperature (if the
parameter control is temperature). In the \emph{coexistence region}
where the two phases have comparable free energies, the distribution
of the order parameter is bimodal. For this reason the Binder
parameter as a function of the temperature also has a characteristic
behaviour, pointed out first by Binder~\cite{Binder84}: it is a
non-monotonic reversed bell behaviour, with a maximal deviation from
Gaussianity in the coexistence region, where the distribution is
bimodal. Indeed, in order to be completely sure that
the deviation from Gaussianity is due to bimodality and not to
other reasons, we also measured the bimodality parameter $b$ defined as
\be
b = \frac{\gamma^2+1}{\kappa+\frac{3(n-1)^2}{(n-2)(n-3)}}.
\label{eq:bimodality}
\ee
where $n$ is the number of data composing the histogram of the
probability distribution, $\kappa$ is the curtosis of the
distribution, i.e. the same ratio of fourth and second order momentum
appearing in the definition of the Binder parameter---see
Eq.~(\ref{eq:Binder})--- and $\gamma$ is the skewness, defined as
\be
\gamma = \frac{\overline{(\Delta n_{\text{eff}})^3}}{\left[\overline{(\Delta n_{\text{eff}})^2}\right]^{3/2}}.
\ee
Indeed, what we find is precisely what Binder presents as the
phenomenology of a first-order transition: the value $\mP_c(N)$ at
which $\mB$ is mostly far from a Gaussian corresponds to the peak of
the bimodality indicator $b$, as well as to the breaking point of
$\overline{n}_{\text{eff}}$.  Eventually, in panel $d)$ of
Fig.~\ref{fig2} we compare the spectral profiles obtained for pumping
rates either well below or well above the lasing threshold.  The
evidence that $\overline{n}_{\text{eff}}$ behaves as the order
parameter is very important for the purpose of the whole discussion.
The main point of presenting here data on equipartition breaking is to
compare them with the standard way to detect ergodicity breaking in
disordered system, i.e. studying the Parisi's order parameter. As we
are going to show, the glass transition that we detect for our system
is a first-order one with respect to its order parameter. It is
therefore very important the information that also the ``order
parameter'' for equipartition breaking behaves as the parameter of a
first-order transition: this puts on solids basis the whole scenario
we are proposing, namely that we are looking at the same underlying
phenomenon from different perspectives. The goal of next section is to
show that the distribution of the equilibrium overlap between
\emph{replicas} confirms this picture.

\begin{figure*}[t!]

  \includegraphics[width=0.9\columnwidth]{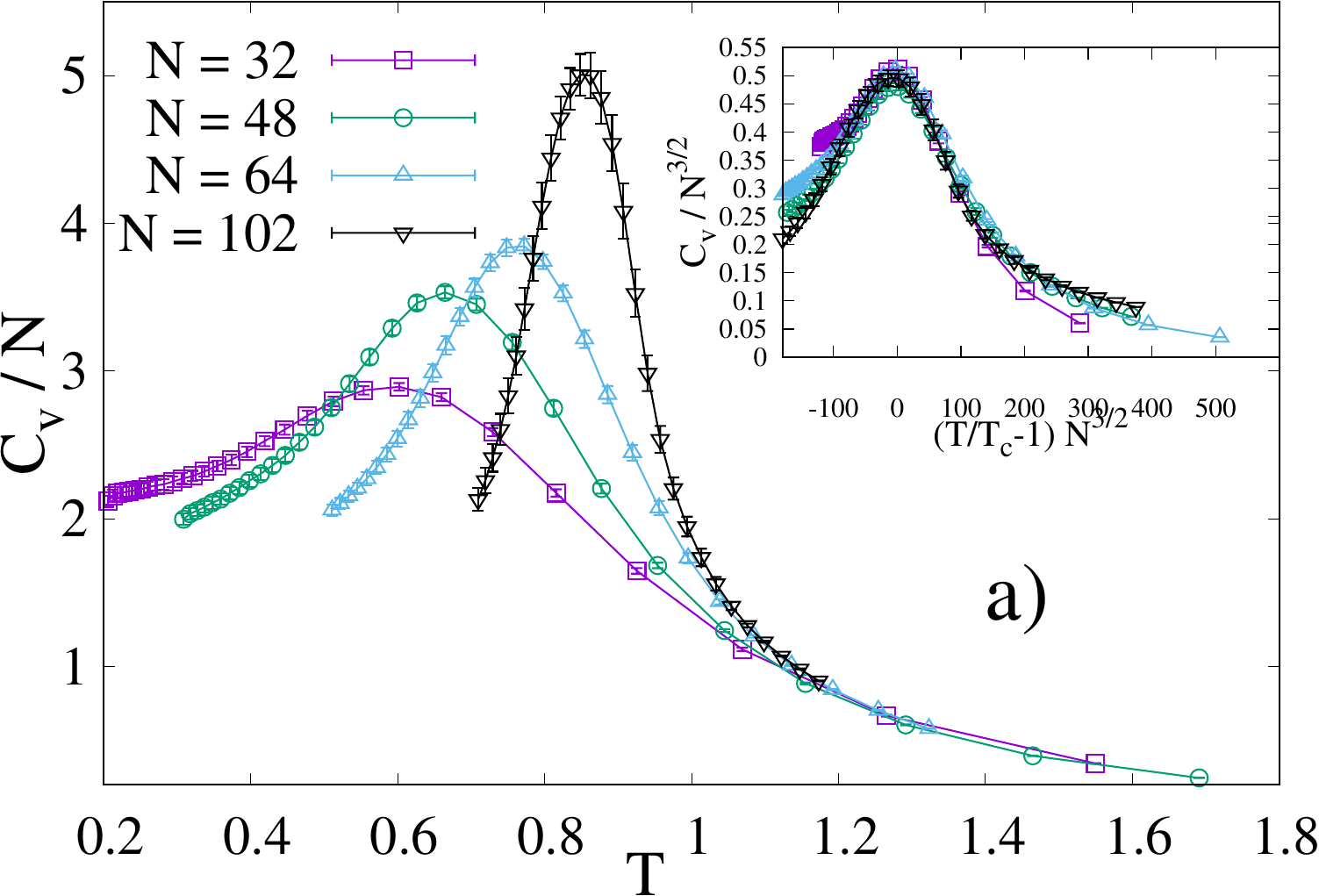}
  \includegraphics[width=0.9\columnwidth]{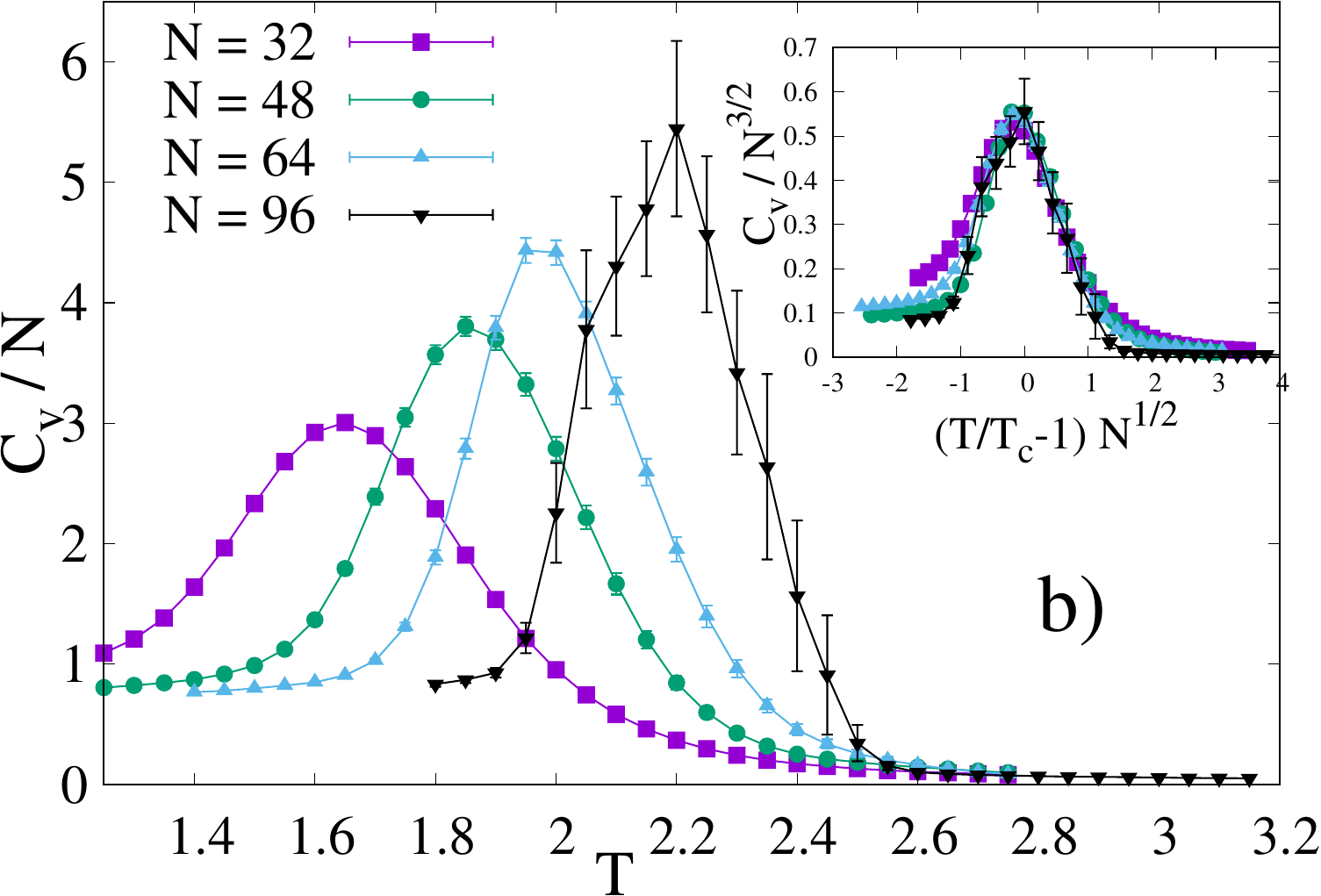}
  \includegraphics[width=0.9\columnwidth]{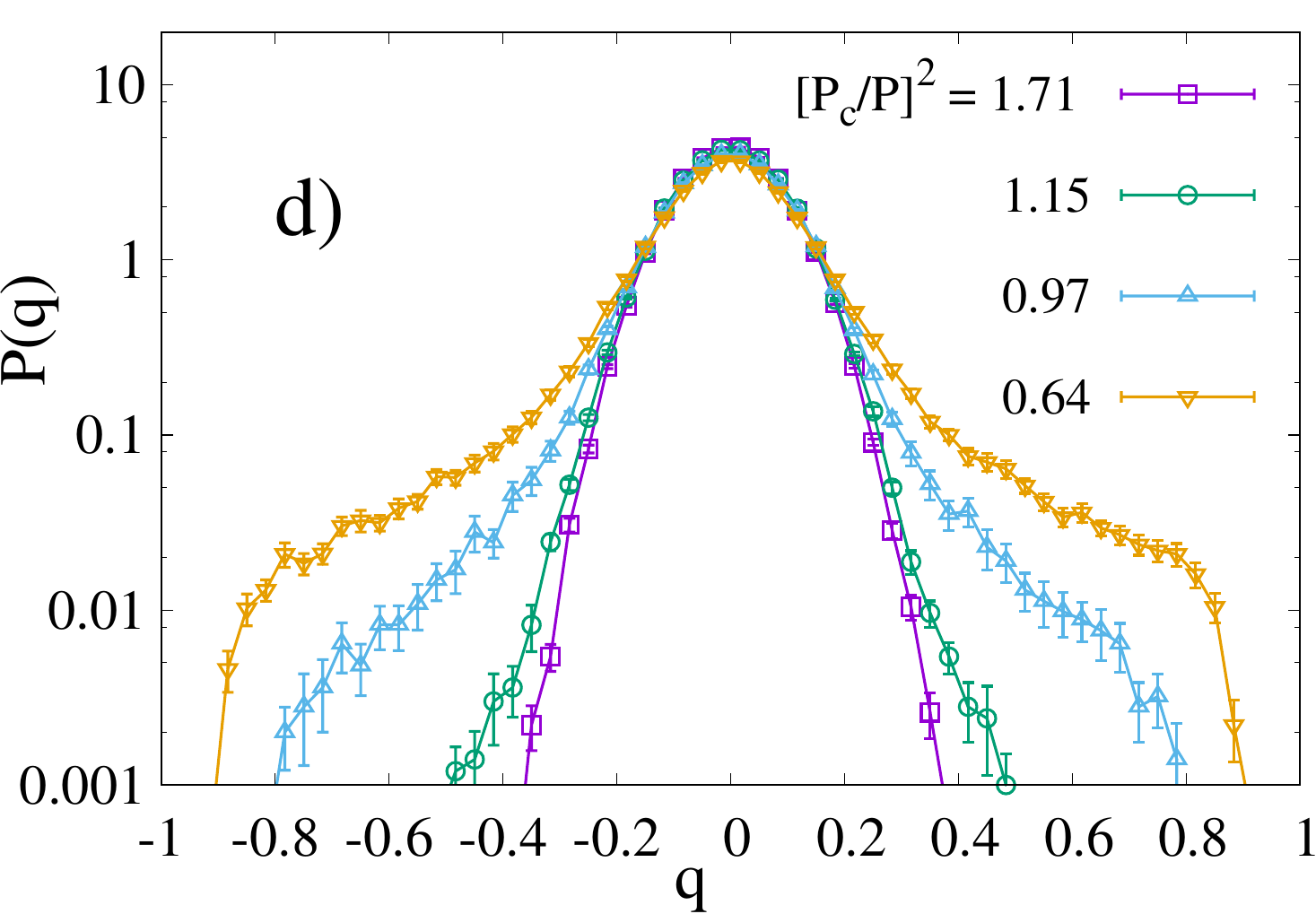}
  \includegraphics[width=0.9\columnwidth]{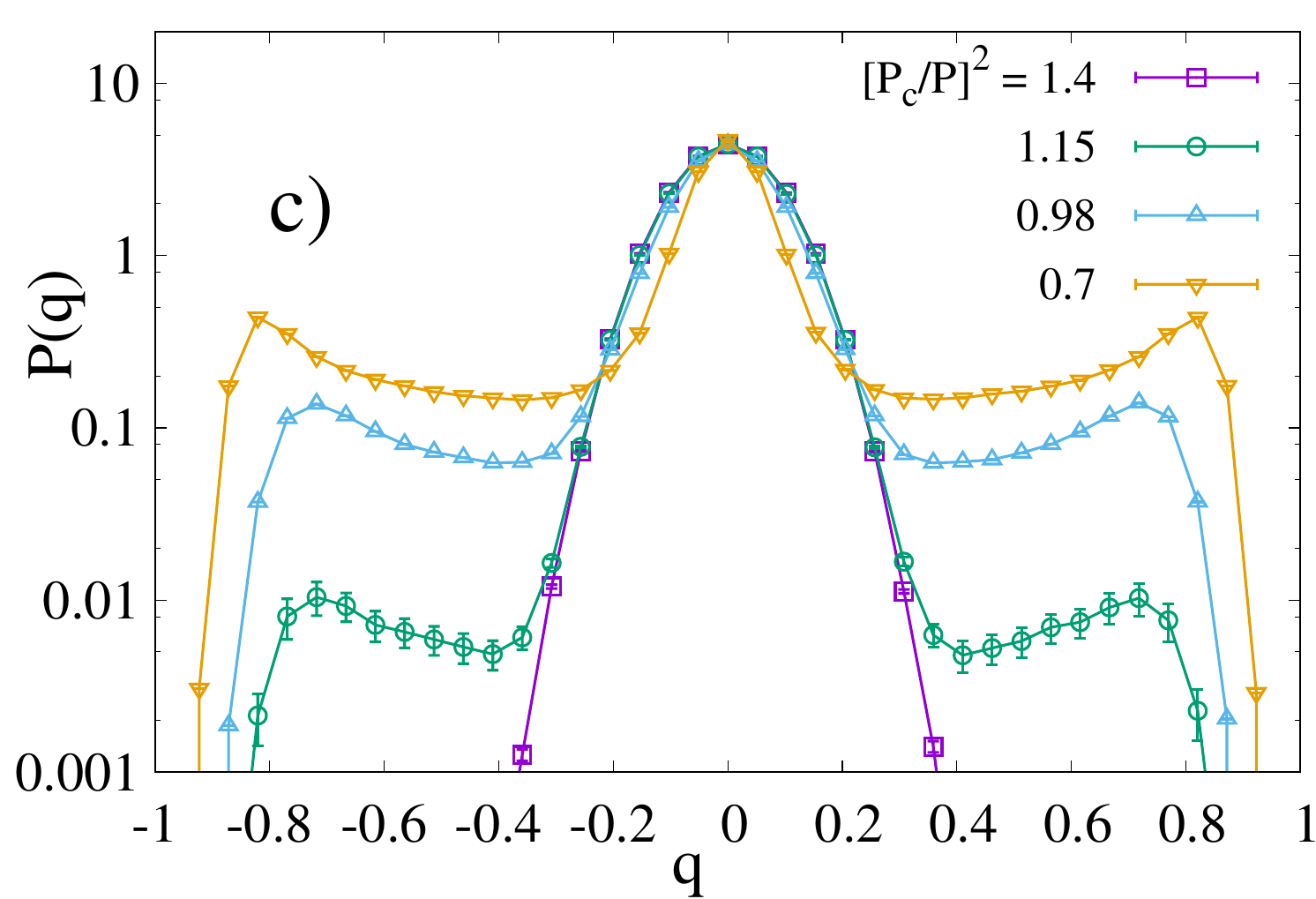}
  \caption{{\it\bf Panel a)} and {\it \bf Panel b)}: Specific heat
    $C_V(T) = \overline{\langle E^2\rangle - \langle E \rangle^2}/T^2$
    as a function of $T$; different curves represent different sizes of
    the system. {\it \bf Panel a)}: 4-phasor model on the mode-locked
    graph.  \underline{{\it Inset}}: Specific heat as a function of
    $\tau N^{3/2}$, where $\tau=T/T_c(N)-1$, curve collapse in the
    scaling region. The four sizes are $N=32,48,64,102$. {\it \bf Panel
      b)}: 4-phasor model on the randomly diluted graph. \underline{{\it
        Inset}}: Specific heat as a function of $\tau N^{1/2}$, curve
    collapse in the scaling region. The four sizes are
    $N=32,48,64,96$. {\it \bf Panel c)}: Random Dilution. Normal modes
    overlap probability distribution for system size $N=64$ at
    temperatures $T/T_c(N)= 1.71, 1.15, 0.98, 0.7$. {\it \bf Panel d)}:
    mode-locked graph. Normal modes overlap probability distribution for
    $N=64$ at temperatures $T/T_c= 1.71, 1.15, 0.97, 0.64$.}
  \label{fig3}
\end{figure*}

\section{Glass transition}
\label{glass-transition}

The so-called Random First-Order Transition
(RFOT)~\cite{Derrida81,Kirkpatrick89,Lubchenko12} is a mixed-order
ergodicity-breaking transition characterized by a specific heat
anomaly and a discontinuity in the order parameter. The latter is
identified as the probability distribution of the overlap $q$ between
equilibrium configurations at a given temperature~\cite{Mezard87}.  In
structural glasses, across the glass transition the distribution
$P(q)$ passes from a unimodal distribution at high temperatures, where
there is a unique state, to a bimodal one with a secondary peak
(actually two symmetric peaks) at low temperatures, because of states
fragmentation into nonequivalent clusters ~\cite{Mezard87}.  This is
the typical behaviour when phase space splits in disjoint ergodic
components: configurations inside the same ergodic component have
typical overlap $q_1$, the overlap between configurations in disjoint
ergodic components being $q_0<q_1$. 

{\color{black}{ Let us stress that this kind of ergodicity breaking is
    not the so called ``spin glass'' transition of the
    Sherrington-Kirkpatrick model \cite{Sherrington75}, where the
    probability distribution of the overlap $P(q)$ is expected to take
    a nontrivial shape, different from a bimodal one, even for
    $N\to\infty$.  In the RFOT case, due to finite size effects,
    $P(q)$ appears continuous but it tends to a bimodal distribution
    in the thermodynamic limit, as is expected for the glass
    transition in the $p$-spin model with $p>2$. The difference is the
    phase transition is well accounted for by the specific heat
    behavior. $C_v$ displays a peak, divergent with $N$, in our
    simulations, whereas in the properly defined spin-glass no
    specific heat singularity is expected to occur at the spin-glass
    transition. }}
   
 We show now that the $4$-phasor random laser model displays these
 RFOT features at the critical pumping power.  We stress that

%
%

Let us start from the specific heat anomaly. In panels $a)$ and $b)$
of Fig.~\ref{fig3} are shown the specific heat $C_V$ curves for
different sizes of the system, where $C_V$ per phasor is
\be \nonumber
\frac{C_V}{N}=\frac{\overline{\langle E^2\rangle -\langle E
    \rangle^2}}{N~T^2}
\ee
The $\langle \phantom{\sharp} \rangle$ and
$\overline{{\phantom{\sharp\sharp}}}$ represent, respectively, thermal
average and average over quenched disorder.  In order to have a term
of comparison, in Fig.~\ref{fig3} we also show the results for the
model (called here random diluted model) that has the same number of
modes, the same distribution of coupling values and with the same
total number of interactions, yet whose set of interacting modes are
not chosen according to Eq. (\ref{eq:FMC_indici}) but, rather,
uniformly randomly selected from the set of all quadruplettes (see
Methods for details).  At all sizes $N$, $C_V/N$ has a characteristic
non-monotonic behaviour, with the position of the peak depending on
$N$. We identify this point as a finite size ``critical temperature"
$T_{c}(N)=1/\mP_c^2$.


The good collapse of the curves at different $N$, shown in the inset
of both panel $a)$ and $b)$ of Fig.~\ref{fig3}, demonstrates critical
scaling, a typical feature of second order phase transitions. The
width of the $C_V$ scaling region for the model with random dilution
is $\tau \sim N^{-1/2}$, where $\tau = T/T_c - 1$: the scaling
exponent $1/\nu=1/2$ is the same predicted by the simplest mean-field
theory for a second order phase transition and is also in agreement
with the reference mean-field model of the RFOT~\cite{Derrida81}. On
the contrary, the scaling region for our 4-phasor mode-locked model
follows the behaviour $\tau\sim N^{-3/2}$, with an exponent
$1/\nu=3/2$, not compatible with a mean-field theory of second-order
transition. This unexpected exponent $3/2$ is probably due to the
correlations in the interaction network induced by the frequency
matching condition, Eq.~(\ref{eq:FMC_indici}). Though it might be a
{\it pre-asymptotic} effect due to too small finite sizes of the
simulated systems, with the present data we cannot rule out the
possibility that it is a behavior persistent even in the thermodynamic
limit.

In order to detect the breaking of ergodicity, we then studied the
overlap in the numerical simulations, where it can be measured as
the similarity between two mode configurations evolving at equilibrium
at the same temperature and with the same realization of random
couplings: {\em two replicas}~\cite{Mezard87}.  Labeling
with the Greek indices $\alpha$ and $\gamma$ two replicas, we have
initially measured (cf. Methods) the phasor overlap $q^{\alpha\gamma}$
\begin{eqnarray}
q^{\alpha\gamma}=\frac{1}{N}\sum_{k=1}^N \bar a_k^\alpha
a_k^\gamma=\frac{1}{N}\sum_{k=1}^N A_k^\alpha
A_k^\gamma\cos(\phi_k^\alpha-\phi_k^\gamma).
\label{eq:Parisi}
\end{eqnarray}
The Greek indices in Eq.~(\ref{eq:Parisi}) denote different replicas,
i.e., {\it independent} configurations at equilibrium at the same
temperature $T$.\\

The Parisi's overlap is characterized by a low temperature non-trivial
distribution in presence of a glass phase~\cite{Mezard87}. In panels
$c)$ and $d)$ of Fig.~\ref{fig3} the phasor overlap distribution
$P(q)$ is shown, respectively for the randomly diluted 4-phasor model
[panel $c)$] and the mode-locked model [panel $d)$] at different
values of $\mP$. In both cases, in the low-$\mP$ ergodic phase, it
turns out to be a symmetric Gaussian.  Then, for $\mP>\mP_{c}$, in the
case of random bond dilution one finds secondary peaks at a finite
distance from the origin, signaling a glassy broken-ergodicity
phase. In the mode-locked graph, actually, this effect is not very
pronounced and shoulders are displayed at the simulated sizes and
powers, rather than proper side peaks. As for the specific heat,
finite size effects turn out to be stronger in the mode-locking graph
than in the randomly bond diluted one.\\

\begin{figure*}
  \includegraphics[width=0.9\columnwidth]{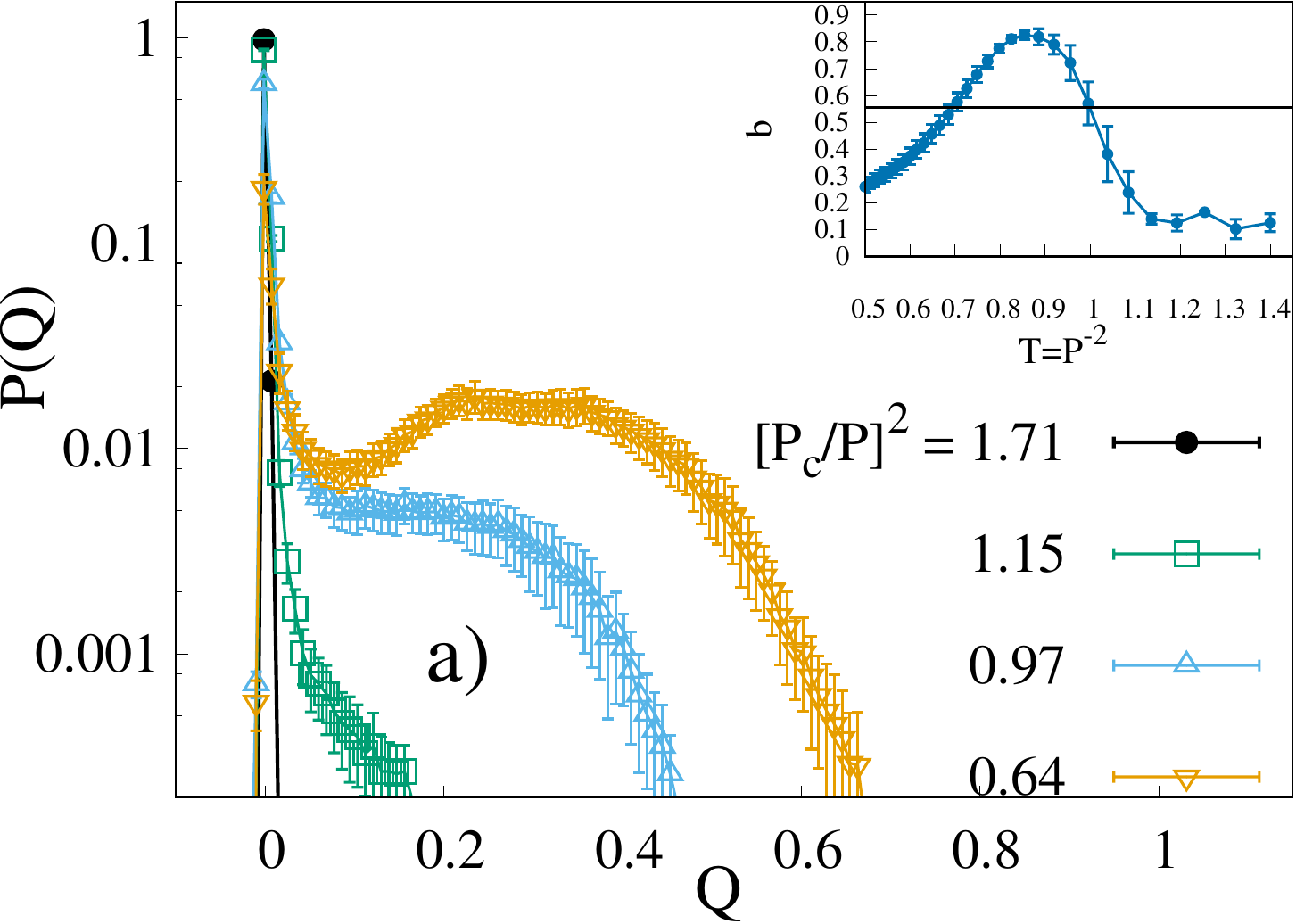}
  \includegraphics[width=0.9\columnwidth]{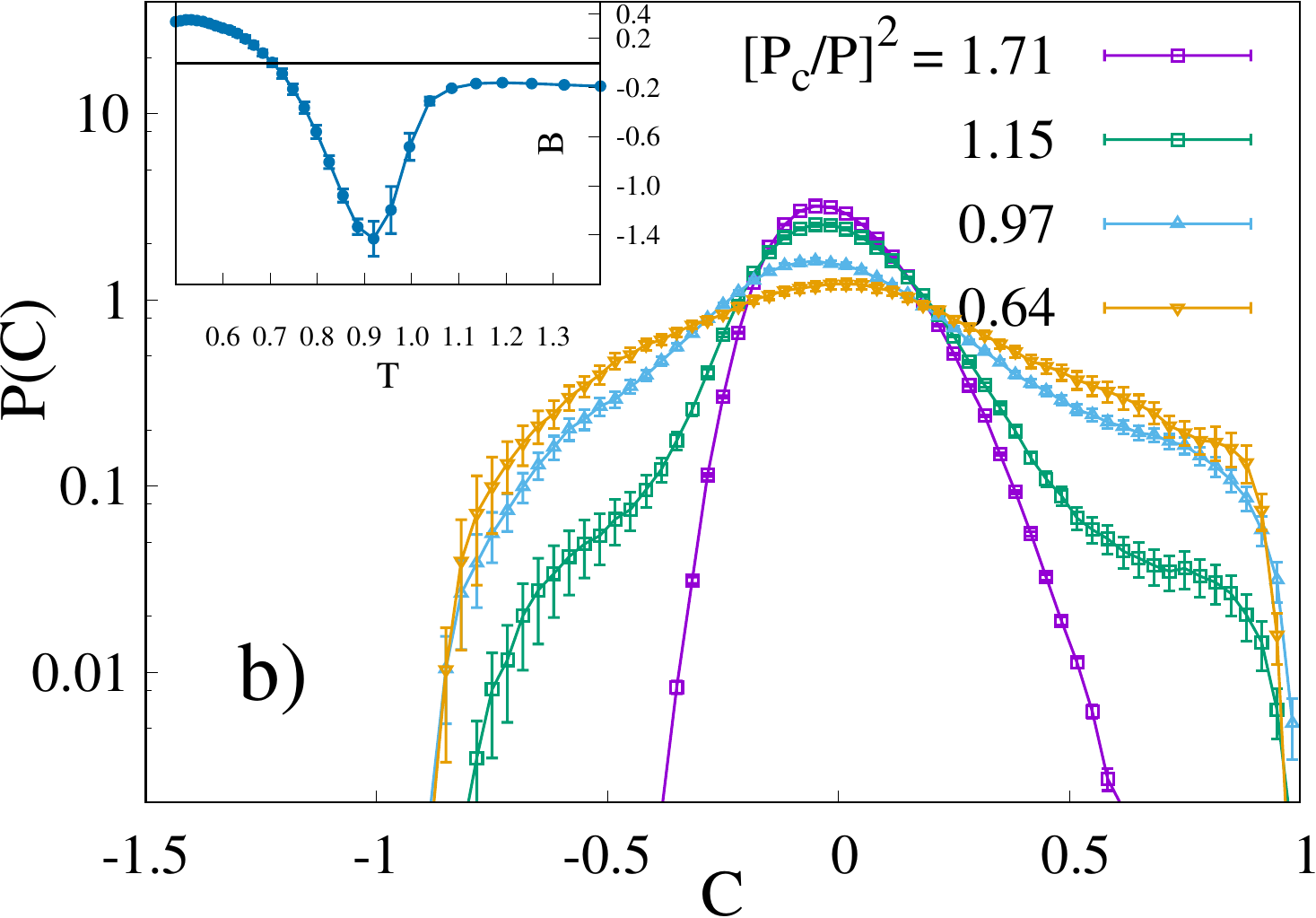}
  \caption{ {\it \bf Panel a)} Plaquettes overlap distribution
    $\mP(\mQ)$ for $N=64$ and $N_4=2^{14}$.  \underline{{\it Inset}}:
    Multimodality parameter $b$ measured for $\mP(\mQ)$, values above
    the threshold $b^*=5/9$ (full black line) indicate a bimodal
    distribution.  {\it \bf Panel b)}: Intensity Fluctuation Overlap
    (IFO) probability distribution $P(\mC)$ for system size $N=64$
    ($N_p=2^{14}$), four different values of the pumping rates:
    $(\mP_c/\mP)^2= 1.71, 1.15, 0.97, 0.64$. \underline{{\it Inset}}:
    Binder parameter $\mB$ measured for $P(\mC)$ as a function of
    $\mP^{-2}=T$; the behaviour is the typical one of first-order
    transitions, with the transition at the minimum of $\mB$.}
  \label{fig4}
\end{figure*}

We further considered another observable, the quadruplettes overlap
probability distribution $\mP(\mQ)$ (see Methods).  The {\it
  quadruplette} overlap $\mQ^{\alpha\beta}$ between the energy stored
in the same set of four coupled modes
\bea
\mathcal{E}_{\vec i} &=& \bar a_{i_1} a_{i_2} a_{i_3} \bar a_{i_4} + \mbox{ c.c. } = \nonumber \\
&=& A_{i_1} A_{i_2} A_{i_3} A_{i_4}~\cos(\phi_{i_2}-\phi_{i_1}+\phi_{i_3}-\phi_{i_4})
\eea
at equilibrium in state $\alpha$ and in state $\gamma$ reads
\be
\mQ^{\alpha\gamma}=\frac{1}{N_4}\sum_{\kappa=1}^{N_4} \mathcal{E}^{\alpha}_\kappa \mathcal{E}^{\gamma}_\kappa,
\label{eq:Parisi-overlap}  \ , 
\ee
where $\kappa$ runs over the ordered list $\{\vec \i\}$ of all $N_4$
non-zero four-body couplings.
While the phasor overlap is computed as the average over $N$
correlated random variables, i.e. the local phasors overlaps $\bar
a_k^\alpha a_k^\gamma$, the quadruplette overlap
$\mQ_\mu^{\alpha\gamma}$ is the average of $N_4=\mO(N^3)$
variables. We, hence, expect it to be less plagued by finite-size
effects. This is, indeed, the case, as shown in Fig.~\ref{fig4}. While
at low pumping rate $\mP$ the quadruplettes overlap has a very peaked
distribution at $\mQ\simeq 0$, for $\mP\approx\mP_c$ we find a clear
signature of a secondary peak at $\mQ> 0$. In the inset of panel $a)$
of Fig.~\ref{fig4} the corresponding multimodality parameter $b$. 

The definition of $b$ is the same given in Eq.~(\ref{eq:bimodality})
above, where skewness $\gamma$ and curtosis $\kappa$ are now defined
as follows:
\bea
\kappa &=& \frac{\overline{\langle (\Delta q)^4 \rangle}}{\left[\overline{\langle (\Delta q)^2 \rangle}\right]^2}, \nonumber \\ 
\gamma &=& \frac{\overline{\langle (\Delta  q)^3 \rangle}}{\left[\overline{\langle (\Delta q)^2 \rangle}\right]^{3/2}},
\label{eq:skew-curt}
\eea
with $\Delta q = q - \overline{\langle q \rangle}$. We find that the
region where the parameter $b$ signals a bimodal distribution of the
overlap is precisely the interval of pumping rates around $\mP_c$. For
the study of the plaquette overlap $\mQ$ distribution we could not use
the Binder parameter as a good indicator because, although clearly
bimodal at the transition, the distribution is not Gaussian far from
the transition.\\


We have presented the numerical evidence that in a statistical
mechanics model for random lasers there is a RFOT glass transition
concomitant with the breaking of equipartition of energy among the
modes.  One question now is to which extent this picture can be
assessed even in experiments. The available technology for the
measurements of light mode phases~\cite{Trebino94,Iaconis98} applies
only to high-power directional impulses: unfortunately this is not the
operating regime of standard random
lasers~\cite{Wiersma08,Leonetti13a}. We cannot therefore rely on
observables which require the measurement of phases. The phasor and
quadruplette overlaps defined in Eqns.~(\ref{eq:Parisi})
and~(\ref{eq:Parisi-overlap}) depend on phases, so they are not
quantities that can be measured in current experiments.  A further
overlap $\mC_{\alpha\beta}$ between intensity fluctuations
$\Delta_k\propto \mI_k-\langle \mI_k\rangle$ can, then, be introduced:
\begin{eqnarray}
\mC^{\alpha\beta} &=&
\frac{1}{N}\sum_{k=1}^N  \Delta_k^\alpha \Delta_k^\beta \nonumber \\ 
\Delta_k^\alpha &\equiv& \frac{\mI_k^\alpha-\langle \mI_k^\alpha\rangle}{2\sqrt{2}\epsilon},
\label{eq:IFO} 
\end{eqnarray}
where $\mI_k=A_k^2$. At the mean-field narrow-band level the Intensity
Fluctuation Overlap (IFO) $\mC_{\alpha\beta}$ is proved to be in a
one-to-one correspondence with the standard overlap, i.e.,
$\mC_{\alpha\gamma} \propto q_{\alpha\gamma}^2$
$\forall~\alpha,\gamma$~\cite{Antenucci15f}. The advantage of
$\mC_{\alpha\beta}$ with respect to $q_{\alpha\beta}$ is that it can
be measured in real random lasers
\cite{Ghofraniha15,Gomes16,Pincheira16, Supratim16,
  Tommasi16,Lopez18,Tommasi18}.
In panel $b)$ of Fig.~\ref{fig4} the distribution $\mP(\mC)$  is shown
for $N=64$ at four different values of the pumping rate $\mP$, two
above and two below the critical $\mP_c$, as determined from the
caloric curve (Fig. \ref{fig2}). At first sight there is no clear
evidence of secondary peaks at high pumping rates for this system
size, although non-Gaussian tails appear in the vicinity of the
transition. It is the study of the Binder parameter $\mB$ dependence
on $\mP$ which reveals how $\mP(\mC)$ brings the signature of a
first-order transition: data are shown in the inset of panel $b)$ of
Fig.~\ref{fig4}. The behaviour of $\mB$ as a function of $\mP$
(plotted as $\mB$ vs $\mP^{-2}=T$ to have a clearer term of comparison
with the literature) is the one characteristic of first-order
transitions~\cite{Binder84}. This behaviour of the IFO order parameter
is, therefore, also consistent with the RFOT scenario and with the
study of the breaking of equipartition in the spectrum: in all the
three cases the ergodicity-breaking parameter behaves as the order
parameter of a first-order transition.

\section{Discussion}
\label{discussion}

By means of Monte Carlo numerical simulations we have analyzed a
statistical mechanical model of the non-linear interactions of light
in a random medium under external optical pumping. We have shown that
the model reproduces the narrowing of the emission spectrum
characterizing the onset of lasing in experiments, see
Fig.~\ref{fig1}, and how it signals an ergodicity-breaking transition.
Indeed, the onset of the random lasing regime from a incoherent regime
of fluorescence displays the properties of a glassy phase transition,
characterized by a diverging specific-heat (typical of continuous
transitions) and a discontinuous order parameter, the overlap, a
feature of first-order transitions. This is termed Random First-Order
Transition in the literature~\cite{Kirkpatrick89}.

Further on, thanks to the quite general properties of the model
analyzed, we were able to show unprecedented evidence of a deep
connection between lack of equipartition - typical of ergodicity
breaking in ordered systems with non-linear
interactions~\cite{Fermi55,Livi85,Chaos05} - and the breaking of
ergodicity as described within the paradigm of
replica-symmetry-breaking in disordered systems~\cite{Mezard87}.

The mode-locked $4$-phasor model is the first example of a system
where ergodicity breaking manifest itself at the same time as a
breaking of the symmetries between replicas and as a lack of
equipartition. \newcolor{We stress that the possibility to reveal this
  new phenomenon is due to the following two salient features of the
  model: the $\mO(N)$ dilution of the non-linear couplings with
  respect to the corresponding mean-field model of the p-spin class,
  induced by the Frequency Matching Conditionand, and the
  local unboundedness of the light-modes amplitudes. While the rule
  for the dilution of couplings is specific of random lasers, it is
  dilution itself which allowed us to see equipartition breaking. The
  model of random laser presented here it is thus, in our opinion,
  only a specific instance of a broader class of systems characterized
  by the concomitance of equipartition and replica symmetry breaking:
  we expect a similar phenomenon to take place for all p-spin models
  on inhomogeneous networks and with locally unbounded variables.}
This important result suggest a possible way to overcome the intrinsic
difficulty usually encountered in the measure of $q$, which is not a
single-experiment observable. The standard protocol is that one needs
to compare the results of several experiments done on the same sample
or of several numerical simulations with the same realization of
quenched disorder to obtain a measure $P(q)$.  Using the jergon of
disordered systems one needs more replicas of the same system.  As we
have shown, the occurrence of a non-trivial $P(q)$ is simultaneous
with the loss of spectral equipartition, and one can, thus, simply
take advantage of latter to detect the ergodicity breaking glass
transition.

In conclusion, as weird as it can sound, the results presented in this
work reinforce the notion that light in random media really looks like
a glassy system and offers, among all physical systems, a good
benchmark to test the existence and the hidden nature of a glass
transition and its connection to nonlinearity.

\section{Ackowledgements}

The authors thank D. Ancora, G. Benettin, L. Biferale, A. Crisanti,
G. Parisi, A. Ponno and A. Vulpiani for useful discussions.  The
research leading to these results has received funding from the
Italian Ministry of Education, University and Research under the
PRIN2015 program, grant code 2015K7KK8L-005 and the European Research
Council (ERC) under the European Union's Horizon 2020 research and
innovation program, project LoTGlasSy, Grant Agreement No. 694925.
G.G. acknowledges the financial support of the Simons Foundation
(Grant No.~454949, Giorgio Parisi).

\section{Appendix}

\subsection{How to build the Mode-Locked graph}

The first step of the numerical study is the generation of the
mode-locked graph with disordered couplings. Our goal is to study what
happens beyond the narrow-band mean-field approximation
of~\cite{Angelani06a,Conti11,Antenucci15a,Antenucci15f}, in particular
when the non-linear interactions is the one of
Eq.~(\ref{eq:hamiltonian}). Operatively, it is easier to describe the
structure of the {\it mode-locked} interaction network as a bipartite
graph where {\it interaction nodes} $\mJ_\mu$ labeled by greek letters
are connected to {\it variable nodes} $A_k$ labeled with latin
letters. Since we have a four-body interaction each interaction node
is always attached to $4$ variable nodes and is defined by the ordered
list of their indices $\mJ_\mu(i,j,k,l)$. The order is relevant
because from the point of view of the energy stored in the interaction
(see Eq.~\ref{eq:hamiltonian}) there are non-equivalent
permutations. The steps to generate the mode-locked graph are as
follows:
\begin{enumerate}
 \item a virtual {\it complete graph} with $\binom{N}{4}$
 interaction nodes is generated; 
\item for each interaction node the three non equivalent index permutations in Eq.~(\ref{eq:hamiltonian}) should be considered:
 $\mP_\mu(i_1,i_2,i_3,i_4)$, 
$\mP_\mu(i_2,i_1,i_3,i_4)$ and $\mP_\mu(i_1,i_2,i_4,i_3)$, even though, for the index order $i_1<i_2<i_3<i_4$ only $\mP_\mu(i_2,i_1,i_3,i_4)$ can be satisfied at most; each time that the mode frequencies (or indeces) 
satisfy the condition (\ref{eq:fmc}),  the corresponding
interaction of the virtual graph is added to the real graph. 
\item The procedure at point $2$ is repeated until a preassigned
  number of interactions in the complete graph is reached; for
  computational reasons this number is the largest power of $2$ below the total number of possible couplings satisfying (\ref{eq:fmc}).
\end{enumerate}
For large $N$, the above procedure tends to cut $O(N)$ of all
interacting quadruplets~\cite{Marruzzo18}. Operatively, for a system
with $N$ complex variables we have drawn a bipartite graph with a
number of interactions scaling as $\mO(N^3)$ and equal to the power of
$2$ soon smaller than the number of all possible interactions
fulfilling the FMC constraint. The number of interaction nodes $N_4$
corresponding to each $N$ is listed in Tab.~\ref{tab1}.\\


Concerning the structure of the topology of the interaction
network it is important to stress that the dimensionality of the
disordered optical medium in real space is scarcely important for the
thermodynamics of the problem: the interaction among modes remains
in any case {\it highly non-local} in the basis of normal modes. The
only quantity which depends on the real-space dimensionality is the
spatial overlap between the normals modes, the information about which
is stored in the disordered coefficients $J_{\vec \i}$ of the
Hamiltonian in Eq.~(\ref{eq:hamiltonian}) as~\cite{Conti11,Antenucci15a}:
\begin{widetext}
\be J_{i_1 i_2 i_3 i_4} = \frac{\imath}{2}
\omega_{i_1}\omega_{i_2}\omega_{i_3}\omega_{i_4}~\int_V d{\bf
  r}~\sum_{\vec \nu}^{x,y,z} \chi_{\vec{\nu}}^{(3)}(\omega_{i_1},\omega_{i_2},\omega_{i_3},\omega_{i_4};{\bf
  r})~ E_{i_1}^{\nu_1}({\bf r})~E_{i_2}^{\nu_2}({\bf
  r})~E_{i_3}^{\nu_3}({\bf r})~E_{i_4}^{\nu_4}({\bf r}). 
\label{def:J}
\ee
\end{widetext}
  where $\chi_{\vec \nu}^{(3)}$ is the non-linear susceptibility of
  the system. We do not consider contributions from the
    first non-linear term $\chi^{(2)}$ in the
    polarization expansion because of the relatively limited band
    width of the emission spectra in random lasers and the consequent
    lack of second harmonic generation, but that term can be easily
    inserted leading to no qualitative difference in the results. Furthermore, we do not explicitly consider linear polarization contributions, that would read
   \be
   \mathcal{H}_2 = -\sum_{k=1}^N a_k\bar a_k g_k \ ,
   \ee
   $g_k = g(\omega_k)$ being the net gain profile (gain taken away losses due to the open cavity) that can be expressed as a function of $\chi^{(1)}(\omega, \bm{r})$.
   In our model though,
    where gain saturation is expressed by the global constraint on the overall intensity distributed in the system 
    \be
    \sum_{k=1}^N a_k\bar a_k = \epsilon N \ ,
    \ee
    and considering for simplicity a constant net gain profile $g_k \simeq g$, this contribution amounts to a constant $\mathcal{H}_2 = - g$, with no role in the dynamics.

In random lasers  couplings as expressed in Eq. (\ref{def:J}) are, in
general, disordered because modes display different spatial
shape and extension \cite{Conti08a,Fallert09}.  The constituents
of the integrals in Eq. (\ref{def:J}) are very difficult
to calculate from first principles. The only specific form of the
non-linear susceptibility has been computed by Lamb
\cite{Lamb64,SargentIII78} for few-modes ordered lasers and no analogue
study for RLs has been performed so far, to our knowledge. 
Integrals like Eq. (\ref{def:J}) in a random medium can be regarded as a sum over
many random variables.  Different couplings
  involving a given mode might, in general, be correlated \cite{Zaitsev10}.
  Since, however, we are interested in the critical behavior, thus in the large size limit 
  of our simulated systems and since
  correlations decay with the size of the system, we adopt as
  working hypothesis a Gaussian  distribution for each
   $J_{\vec{\i}}$:
 \begin{eqnarray}
P(J_{\vec{\i}})= \sqrt{\frac{N^{2}}{2\pi }} \exp\left\{
-\frac{N^{2}J_{\vec{\i}}^2}{2}
 \right\}
\label{def:Gauss}
\end{eqnarray} 
The scaling of  variance with $N$,
  $\langle J^2 \rangle \sim N^{-2}$,  guarantees energy
  extensivity. There are very many couplings but each one is vanishingly small. Macroscopic phenomena like lasing occur because 
  the system undergoes a transition to a collective behavior. Even in presence of randomness cooperativity is the leading mechanism.
We further stress that, from the perspective of probing RFOT, considering correlated $J$'s leads to qualitatively
analogue phase diagram as it is well known in spin-glass systems such us,
e.g., the Random Orthogonal Model \cite{Marinari94a,Parisi95b}.  

If we look at interactions in the space of normal modes the system is
{\it infinite-dimensional}: any degree of freedom participates to an infinite 
($O(N^2)$) number of interactions in the thermodynamic limit $N\to\infty$. That is why we expect the mean-field glass
transition scenario drawn in~\cite{Antenucci15a,Antenucci15f} to be
quite robust for the mode-locked $p$-phasor, even if the narrow-band hypothesis \cite{Gordon02}
 is removed.
  
Last but not least, the non-locality of interactions between light
modes also guarantees that phenomena like energy localization, a {\it
  pathology} of sparse networks~\cite{Antenucci15c,Antenucci15d}, are
avoided.



\subsection{Replicas in numerical simulations}

Replicas are independent equilibrium configurations sampled with the
same quenched disorder. This definition corresponds to the protocol
used in numerical simulation. One ``replica'' of the system is
represented by the {\it swarm} of $N_{PT}$ configurations used for a
given instance of the Parallel Tempering MC dynamics. Different
istances of the PT dynamics characterized by the same set of quenched
couplings $J_{\vec{\i}}$ and the same interaction network between
modes are different replicas.

For $N=32$, $N=48$ and $N=64$ for each disorder instance we simulated
$4$ replicas, which gave us the availability of $6$ independent values
of $q^{\alpha\beta}$: $q^{12}$, $q^{13}$, $q^{14}$, $q^{23}$,
$q^{24}$, $q^{34}$. For $N=102$ we simulated two replicas for each
instance of the disorder.  To accumulate statistics for $P(q)$ we
measured values of $q^{\alpha\beta}$ comparing replicas {\it at the
  same iteration} of the PT dynamics, each $640$ iterations. Since the
distribution $P(q)$ is not self averaging~\cite{Mezard87}, for each
size of the system we have sampled the equilibrium measure for
$N_{\text{sample}}\approx 100$ istances of the disorder.\\

It is useful to clarify also how we measured in practice all thermal
averages indicated with angular brackets, $\langle \mO[{\bf A}]
\rangle$. Once that the system reaches equilibrium at a given temperature/power, at a time in Monte Carlo sweep units of $N_{\rm term} < N_{\text{sweep}}/4$,
 thermal averages were measured as time averages
along the dynamics, along the second half of each run:
\be
\langle \mO[{\bf A}] \rangle = \frac{2}{N_{\text{sweep}}} \sum_{i=N_{\text{sweep}}/2}^{N_{\text{sweep}}}  \mO[{\bf A}_i].
\ee

\section{References}

\bibliographystyle{naturemag}
\bibliography{Lucabib}

\end{document}